\documentclass[preprint,superscriptaddress]{revtex4-1} %% use 12pt for preprint option
\usepackage{amsmath,amssymb}
\usepackage{graphicx}
\usepackage{epsfig}
\usepackage[sort&compress]{natbib}

\begin{document}

\title{Single shot polarimetry imaging of multicore fiber}
\author{Siddharth Sivankutty}
\affiliation{Aix Marseille Universit\'{e}, CNRS, Centrale Marseille, Institut Fresnel, UMR 7249, 13397 Marseille Cedex 20, France}

\author{Esben Ravn Andresen}
\author{G\'{e}raud Bouwmans}
\affiliation{Universit\'{e} Lille 1, Laboratoire PhLAM, 59655 Villeneuve d'Ascq, France}

\author{Thomas G. Brown}
\affiliation{The Institute of Optics, University of Rochester, Rochester, NY 14627, U.S.A.}

\author{Miguel A. Alonso}
\email{malonso@UR.Rochester.edu}
\affiliation{The Institute of Optics, University of Rochester, Rochester, NY 14627, U.S.A.}
\affiliation{Center for Coherence and Quantum Optics, University of Rochester, Rochester, NY 14627, U.S.A.}

\author{Herv\'{e} Rigneault}
\email{herve.rigneault@fresnel.fr}
\affiliation{Aix Marseille Universit\'{e}, CNRS, Centrale Marseille, Institut Fresnel, UMR 7249, 13397 Marseille Cedex 20, France}

\begin{abstract}
We report an experimental test of single-shot polarimetry applied to the problem of real-time monitoring of the output polarization states in each core within a multicore fiber bundle. The technique uses a stress-engineered optical element together with an analyzer and provides a point spread function whose shape unambiguously reveals the polarization state of a point source. We implement this technique to monitor, simultaneously and in real time, the output polarization states of up to 180 single mode fibers cores in both conventional and polarization-maintaining fiber bundles. We demonstrate also that the technique can be used to fully characterize the polarization properties of each individual fiber core including eigen-polarization states, phase delay and diattenuation.
\end{abstract}
\maketitle

Multicore fiber (MCF) are important for a broad range of applications spanning over the fields of telecommunication \cite{Zhu_2010, Van_Uden_2014}, high power fiber lasers \cite{Glas_1998, Li_2007} and medical endoscopes \cite{Gmitro_1993,Laemmel_2004}. Recently there has been a growing interest for coherent beam combining \cite{Bellanger_2008} using MCFs both in the fields of broadband laser amplification \cite{Mansuryan_2012, Ramirez_2015} and lens-free endoscopy \cite{Thompson_2011, Andresen_2013_1, Andresen_2013_2}. In coherent beam combining, an initial laser beam is, spectrally and/or spatially, split into $\mathrm{N}$ beamlets that are launched and propagate into the $\mathrm{N}$ individual cores of a MCF. At the exit of the bundle, the different beamlets are recombined in a single beam and their relative phases are adjusted by a servo for coherent synthesis of an amplified ultrashort pulse or a focused spot, depending the application. Optimal temporal and/or spatial, focus at the bundle output requires to actively manage the relative beamlet phase and group delay \cite{Andresen_2015, Roper_2015} together with their polarization state. Recent works report the use of polarization-maintaining $(\mathrm{PM})$ MCF \cite{Ramirez_2015, Kim_2016} to ensure that the output beamlets exhibit identical polarization states for optimal spectral and/or spatial interference. Nevertheless, accurate core-by-core polarization state characterizations have not yet been carried out, neither the relative stability of these states during a fiber bend or twist. Such monitoring is important for calibration purposes and to infer the polarization origin of output power evolutions during fiber bends. Ideally one would like to monitor in real time the polarization state at the output of 10 to $\mathrm{~}$10000 fiber cores constituting the fiber bundle of interest. This is quite challenging since standard polarimetry imaging techniques usually require time-sequential operations such as, for instance, analyzing the intensity of the light passing through a rotating quarter wave plate and a fixed polarizer \cite{Collet_1992}. 'Single shot' polarimetry imaging has been also reported where the wavefront is divided and directed to different sensors \cite{Azzam_1985}, or where a clever combination of Savart plates is used \cite{Luo_2008}. These 'single shot' operations come with a significant technical complexity.

\begin{figure}[t]
\centerline{\includegraphics[width=0.87\linewidth]{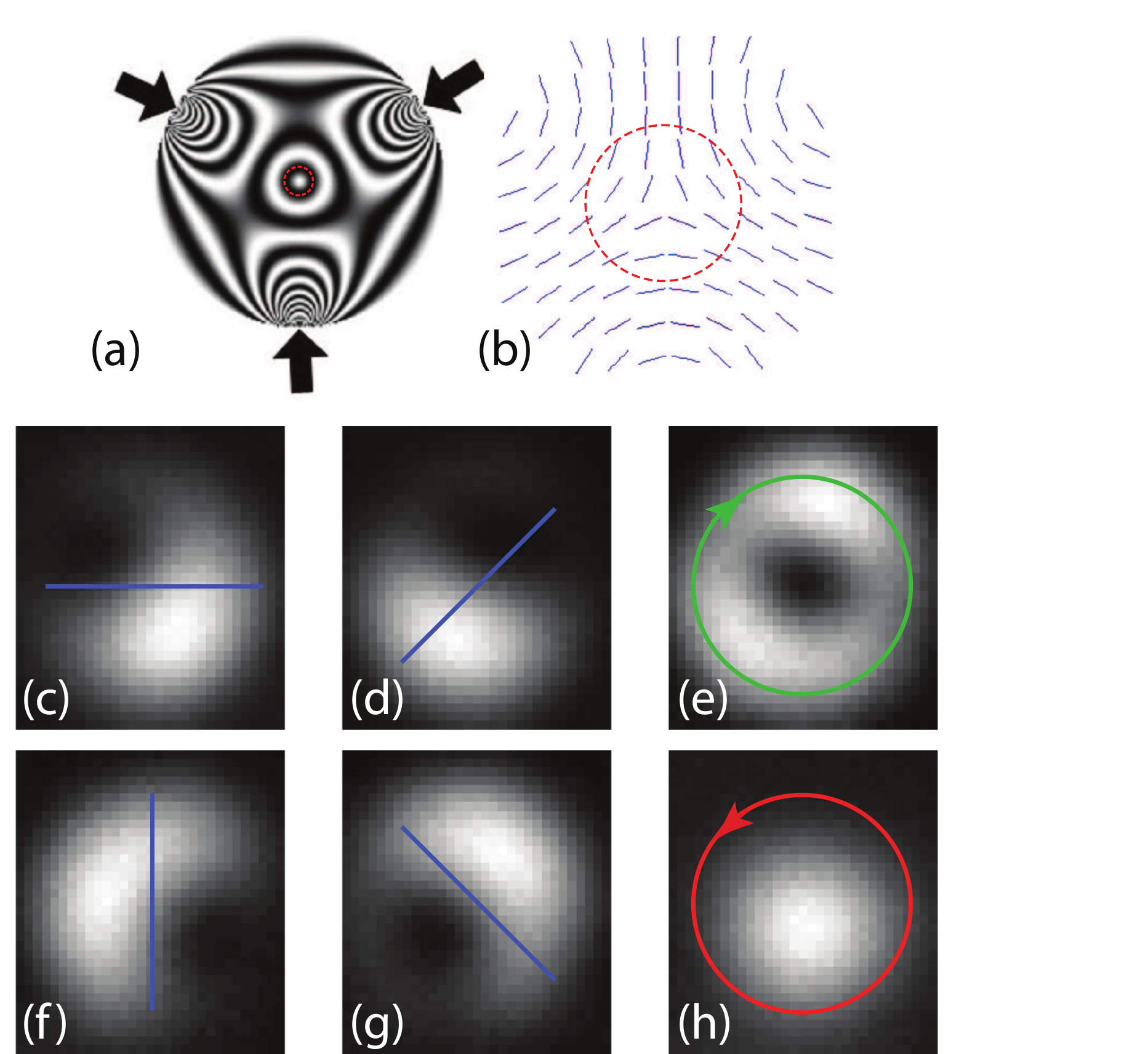}}
	\caption{(a) Simulation of stress-engineered optical element (SEO) and used area (inside red dashed circle), where white (black) indicates that the retardance is an integer (half-integer) multiple of the wavelength. (b) Plot of spatially-varying fast axis orientation near the window's center. (c-h) Measured point spread function (PSF) corresponding to the image of an individual fiber core (approximately a point source) through the combination of SEO and a left-handed circular analyzer (LHCA) (see Fig.~\ref{fig2}(a)), for (c) horizontal, (d) linear at $+45^\circ$, (e) right-hand circular, (f) vertical, (g) linear at $-45^\circ$, (f) left-hand circular polarization states, where a representation of the polarization is overlaid with blue indicating linear polarization, and green and red indicating right- or left-handedness.}
	\label{fig1}
\end{figure} 

A method of polarimetry was recently demonstrated that can infer complete polarization information using single-exposure images of discrete point-like light sources using an optical window subjected to stress with trigonal symmetry \cite{Ramkhalawon_2013}. This stress-engineered optical (SEO) element \cite{Spilman_2007} provides, near the window's center Fig.~\ref{fig1}(a), a retardance that increases linearly with the radius and an orientation of the fast axis that precesses with the azimuth Fig.~\ref{fig1}(b). The SEO can be viewed as a spatially-varying wave plate: each point over the cross section of the test beam experiences a different combination of retardance and orientation. Therefore a collimated beam passing through the desired section of the  SEO (the rest being blocked) acquires an intricate polarization structure \cite{Spilman_2007stress, Beckley2010} that, when observed at the focus of a lens, and after passing through a left-handed circular analyzer (LHCA), provides a point spread function (PSF) whose shape unambiguously reveals the polarization state \cite{Ramkhalawon_2013}. Fig.~\ref{fig1}(c-f) shows measured PSFs at the focal plane for the standard six reference polarization states. As discussed in \cite{Ramkhalawon_2013}, the PSF is a linear function of the Stokes parameter vector, so the state of polarization (including how polarized or unpolarized the light is) can be inferred from the measured PSF in real time, without the need for sequential measurements. Zimmerman {\it et al.} \cite{Zimmerman2014} described how an extension of this principle allows one to capture polarization information from an image sampled by a pinhole array  by placing the SEO in the pupil (fourier conjugate) plane.

In this letter, we report on using SEO polarimetric imaging to monitor in real time the polarization state at the output of a MCF. We concentrate on the polarization state evolution when the fiber bundle is illuminated with rotating linear polarization states and retrieve, for each fiber core, the equivalent phase plate. We also investigate the stability of the output polarization state when the MCF is slightly bent.

\begin{figure}[htbp]
\centerline{\includegraphics[width=\linewidth]{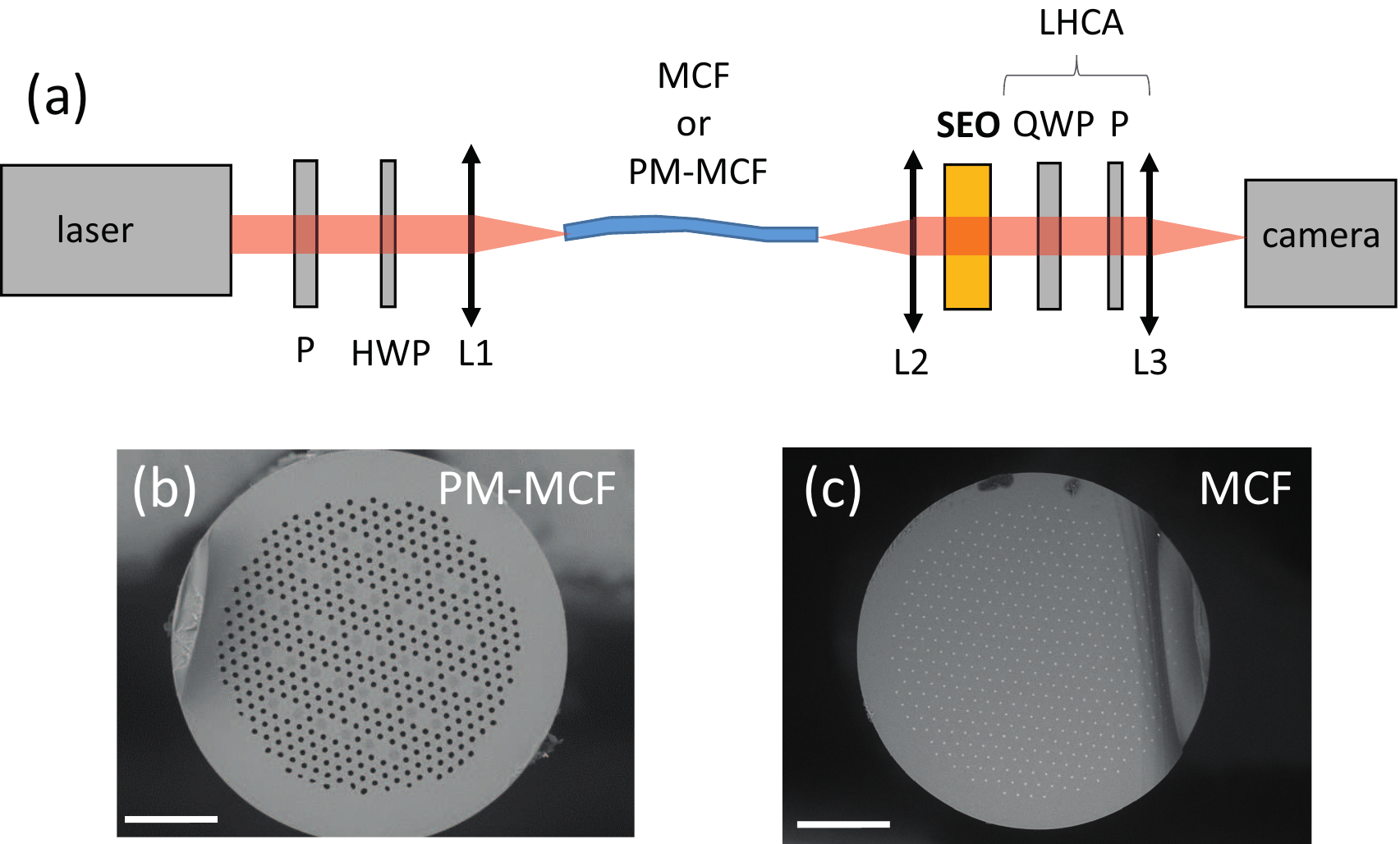}}
	\caption{ (a) Setup: P polarizer, HWP half wave plate, lens $L_{1}$ $\mathrm{f_1}$=25~mm, MCF multi core fiber, lenses $L_{2}$ and $L_{3}$ represent a set of lenses that image the fiber face onto the camera with a magnification of  \textasciitilde 13x, SEO stress-engineered optical element, QWP quarter wave plate, LHCA left-handed circular analyzer (b) 19 cores PM single mode MCF; (c) 475 cores non-PM single mode MCF; scale bar: 100~$\mu$m.}
	\label{fig2}
\end{figure}

Fig.~\ref{fig2}(a) shows a schematic of the optical setup. A pulsed Ytterbium laser (t-pulse, Amplitude systemes, 1030~nm, 50~MHz, 1~W) is controlled in polarization by using a polarizer and a half wave plate. The laser beam is then weakly focused to illuminate the entrance facet of the tested MCF. The output MCF facet is imaged onto a CMOS camera after crossing a SEO element (placed at a fourier conjugate plane) and a left-handed circular analyzer (composed of a quarter wave plate and a polarizer). Two fused silica MCFs are considered in this work (Fig.~\ref{fig2}(b-c)), the first one consisting of 19 polarization-maintaining (PM-MCF) single mode fiber cores (pitch 40~$\mu$m, core diameter 18~$\mu$m, outer diameter 348~$\mu$m, length 50 cm), the second one (MCF) made up of 475 non-PM single mode fiber cores (pitch 15.3~$\mu$m, core diameter 3.1~$\mu$m, outer diameter 421~$\mu$m, length 30 cm). Both MCFs exhibit very low cross talk, below -30~dB and were held straight during the measurements.  

\begin{figure}[h]
\centerline{\includegraphics[width=0.95\linewidth]{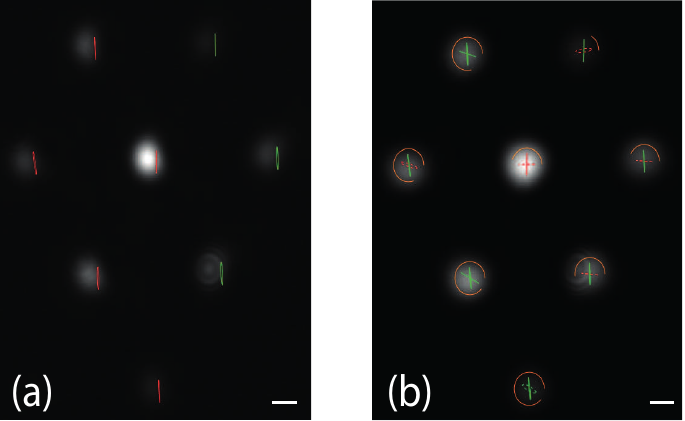}}
	\caption{(a) SEO PSFs (grey scale) and output polarization states when 8 cores of the PM-MCF are illuminated with  vertically polarized light aligned along one of the PM-MCF principal axis. (b) Total (unpolarized) transmitted light (grey scale) and overlaid eigen-polarizations associated to each fiber core, the angle traced by the orange arc corresponds to the phase delay between the two eigen-polarizations (see text). Green represents right-handedness and red left-handedness. Scale bar: 10~$\mu$m.  Also see Visualization 1.} 
	\label{fig3}
\end{figure}

The system is calibrated by taking images with unpolarized illumination of the fiber input (or with the sum of images with two orthogonal illumination polarizations) and where polarizers are used after the fiber to create the six reference PSFs for all cores, such as those shown in Fig.~\ref{fig1}(c-h) for one core. By then performing a linear fit of the PSF for each core with the corresponding reference PSFs, one can determine the Stokes parameters.
Figure~\ref{fig3}(a) shows an image of the SEO PSFs (grey scale) obtained when illuminating 8 cores of the PM-MCF with  linearly polarized light aligned along one of the PM-MCF principal axes. Note that the PM-MCF input facet was illuminated with a quasi-plane wave, without special core coupling optimization. Overlaid on Fig.~\ref{fig3}(a) are the output polarization ellipses (as directly deduced from the SEO recorded PSFs of $57 \times 57$ pixels each). Note that these outputs are all approximately vertically polarized, a result consistent with the aligned linear polarized illumination and the PM characteristics of this MCF.
The movie in Visualization 1 shows the same images for several linear input polarizations (rotating the HWP in Fig.~\ref{fig2}(a)). Clearly, two linear output polarizations are obtained when the PM-MCF is illuminated with linearly polarized input light along two orthogonal directions (close to 0~° and 90~°). Visualization 1 also shows the trajectories over the Poincar\'e sphere of the output polarization states corresponding to the 8 cores. These paths are all approximately great circles (that is paths that draw the largest circle on the Poincar\'e sphere), and this indicates that each core behaves like a phase plate. This phase-plate-like behavior is represented pictorially in Fig.~\ref{fig3}(b), where on each core, we overlay its two eigen-polarizations (solid and dashed lines). The angle traced by the orange arc corresponds to the phase delay between these two eigen-polarizations states and its handedness ( right and left) is represented by the colours green and red. As expected, the two eigen-polarizations are mostly linear and oriented along two orthogonal axis on this PM-MCF.  Note that since the Jones matrices are not perfectly unitary, the eigen polarizations might not be exactly orthogonal. As for the phase delay, it arises from the differential phase between the two eigen-polarization that has been accumulated during the propagation along the fiber length and this is indeed expected to vary from core to core.

\begin{figure}
\centerline{\includegraphics[height=0.9\linewidth]{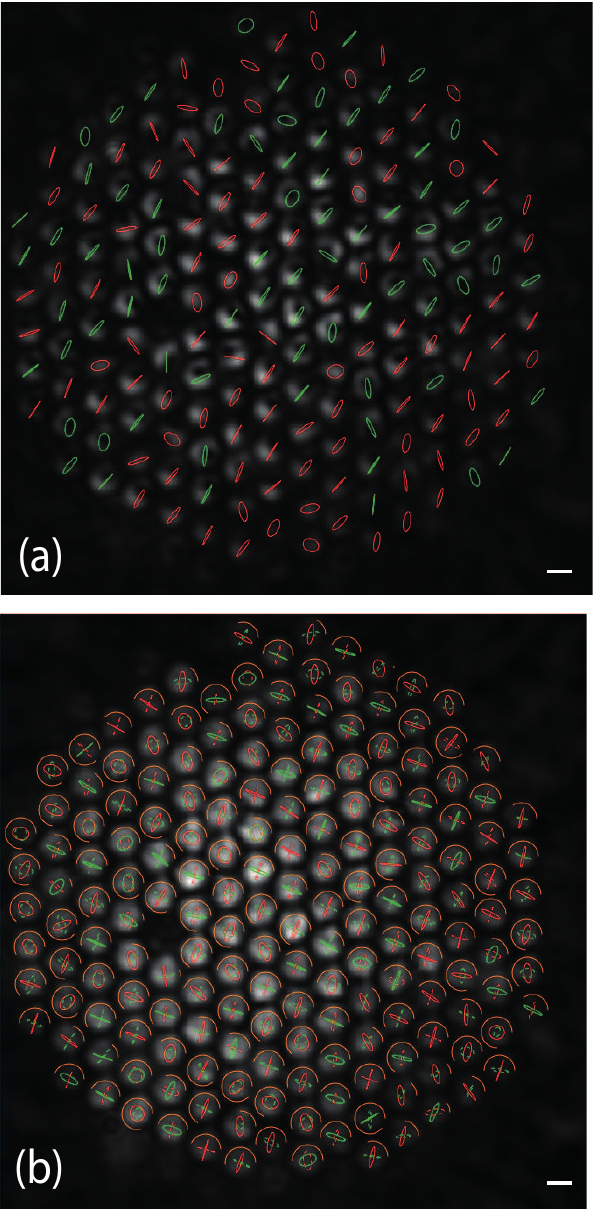}}
	\caption{(a) SEO PSFs (grey scale) and output polarization states when 180 cores of the non PM MCF are illuminated with a vertically polarized light. (b) Total (unpolarized) transmitted light (grey scale) and overlaid eigen-polarizations associated to each fiber core, the angle traced by the orange arc corresponds to the phase delay between the two eigen-polarizations. Green represents right-handedness and red left-handedness. Scale bar: 10~$\mu$m.  Also see Visualization 2.}
	\label{fig4}
\end{figure}

We now move to the non-PM 475 cores MCF, that we denote 'MCF' for simplicity. Here again the MCF facet is illuminated with a quasi-plane wave without special care to optimize core coupling. Figure~\ref{fig4}(a) shows the recorded SEO PSFs (grey scale), ~$31 \times 35$ pixels each, together with the overlaid output polarization states for 180 cores, when the MCF is illuminated with vertically polarized light. As in the previous case, green and red ellipses represent right-handedness and left-handedness, respectively. Contrary to the PM-MCF, there are no apparent correlations in core to core output polarization states.
The movie in Visualization 2 shows the same images when rotating the linear input polarizations, together with the trajectories over the Poincar\'e sphere of the output polarization states corresponding to each core. As previously, the paths follow approximately great circles over the Poincar\'e sphere surface indicating that each core acts as a phase plate. Using this polarimetric information, we represent in Fig.~\ref{fig4}(b) the two eigen-polarizations associated to each core, together with the phase delay (orange arc) between them. As previously, for each core, the underlying unpolarized transmitted intensity is also shown (grey scale). Careful inspection of Fig.~\ref{fig4}(b) reveals correlation between core to core eigen-polarization axis. This residual anisotropy probably comes from the stress present in the MCF preform and is a direct result of its fabrication process. As in the PM-MCF case the phase delay between the two eigen-polarization shows no core to core correlation and comes from the phase accumulated during propagation.

Our SEO-based polarimetric measurement can also address diattenuation in each fiber core, an effect related to the differential propagation loss arising between the two eigen-polarizations. This can be seen in Fig.~\ref{fig3}(b) and Fig.~\ref{fig4}(b) where the measured diattenuation is indicated by the size of the second (dashed) eigen-polarization ellipse relative to the non-dashed one (when diattenuation is absent the sizes of the dashed and non-dashed ellipse are the same). When diattenuation is present, the output polarization state trajectory is no longer a great circle over the Poincar\'e sphere as the input polarization is varied and describes a great circle trajectory over the Poincar\'e sphere. 
The SEO based polarimetry imaging allows then the full characterization of each core as a Jones matrix, whose pictorial representation in terms of eigen-polarizations, diattenuation and relative phase is given in Fig.~\ref{fig3}(b) and Fig.~\ref{fig4}(b).

\begin{figure}
\centerline{\includegraphics[width=0.95\linewidth]{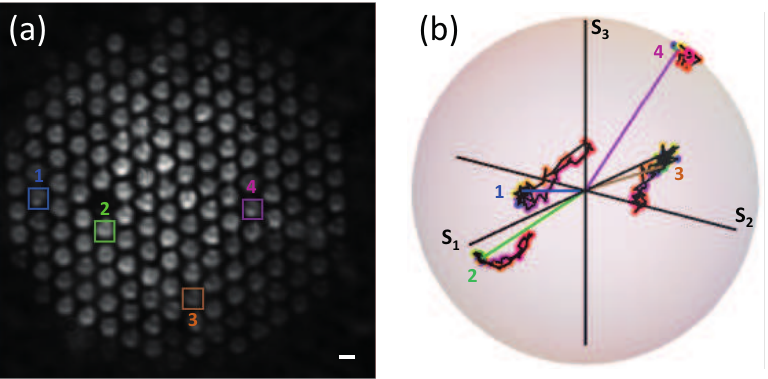}}
	\caption{(a) SEO PSFs (grey scale) of the 180 cores non PM MCF where four cores are selected to display (b) the trajectories of their output polarization state over the Poincar\'e sphere when applying manual wiggling and bending , to guide the eye, time is color coded. Scale bar: 10~$\mu$m. See Visualizations 3-5 for movies of the same.}
	\label{fig5}
\end{figure}

We finally investigate the influence of a slight twist or bend on the fiber core output polarization states. We take here full advantage of our single shot polarimetric measurement by recording video rate images of the SEO PSFs for 180 illuminated cores while bending the fiber. Visualization 3 shows 180 SEO PSFs (grey scale) and the associated output polarization states when the MCF is illuminated with vertically polarized light and manually wiggled and slightly bent (radius of curvature approximately 10~cm over a fiber length of few cm). Visualization 4 shows the applied manual wiggling and bending. Fig.~\ref{fig5}(a) shows the SEO PSFs where we have selected four cores and display their trajectories over the Poincar\'e sphere Fig.~\ref{fig5}(b) while applying manual bending. Visualization 5 shows a movie of the same. Quite surprisingly, the output polarization states are weakly affected by the bend suggesting that the phase delay accumulated along propagation between the two eigen-polarization states remains fairly constant in the case of wiggles and slight bends. This clearly demonstrates that SEO polarimetric imaging can monitor in real time the polarization state at the output of a MCF having a very large number of cores.\\

In conclusion, we have described and tested a  method of single shot polarimetry \cite{Ramkhalawon_2013} applied to the characterization of two different types of multicore fiber (MCF). Using a stress-engineered optical element followed by an analyzer, we can simultaneously retrieve the polarization states of the light emitted by a large number of fiber cores simply by looking at the individual images of the cores at the focus of a lens. The technique is real-time and we have shown that it allows the full polarization characterization of each fiber core. Our approach is interesting for applications such as coherent beam combining in MCFs where the output polarization in each core has to be identical. It also opens new directions in the field of lens-free endoscopy \cite{Thompson_2011, Andresen_2013_2} where performing polarimetric imaging together with full polarization and phase control at the MCF output gives access to the full vector PSF control, in a way similar to what has been done in microscopy \cite{Beversluis_2006,Kenny_2012}. This would open the door for important new avenues in endoscopy including structural molecular and biological imaging \cite{Brasselet_2011}.

\section*{Funding Information}

\textbf{Funding.} We acknowledge financial support from the CNRS Mission Interdisciplinaire DEFI CODIM, Aix-Marseille University A*Midex (noANR-11-IDEX- 0001-02), Agence Nationale de la Recherche (10-INSB-04-01, 11-INSB-0006), ISERM PC201508, and from the National Science Foundation through awards PHY-1068325 and PHY-1203931.

\section*{Acknowledgments}
This work was partially supported by IRCICA USR 3380 Univ. Lille - CNRS

%\textbf{Acknowledgment.} Formal funding declarations should not be included in the acknowledgments but in a Funding Information section as shown above. The acknowledgments may contain information that is not related to funding:

%\section*{References}

\end{document}